\documentclass[sigconf,authorversion]{acmart}
\AtBeginDocument{%
  \providecommand\BibTeX{{%
    \normalfont B\kern-0.5em{\scshape i\kern-0.25em b}\kern-0.8em\TeX}}}

\copyrightyear{2024} 
\acmYear{2024} 
\setcopyright{acmlicensed}\acmConference[WWW '24 Companion]{Companion Proceedings of the ACM Web Conference 2024}{May 13--17, 2024}{Singapore, Singapore}
\acmBooktitle{Companion Proceedings of the ACM Web Conference 2024 (WWW '24 Companion), May 13--17, 2024, Singapore, Singapore}
\acmDOI{10.1145/3589335.3651535}
\acmISBN{979-8-4007-0172-6/24/05}

\usepackage{enumitem}
\usepackage{multirow}
\usepackage{algorithm}
\usepackage{amsmath}
\usepackage{bbm}
\usepackage{balance}
\usepackage{subfigure}
\usepackage{siunitx,booktabs}
\usepackage{marvosym}

\begin{document}

\title{MART: Learning Hierarchical Music Audio Representations \\with Part-Whole Transformer}


\author{Dong Yao$^*$}
\affiliation{%
  \institution{Zhejiang Unversity, China}
  \country{}}
\email{yaodongai@zju.edu.cn}

\author{Jieming Zhu$^*$}
\affiliation{%
  \institution{Huawei Noah's Ark Lab, China}
  \country{}}
\email{jiemingzhu@ieee.org}

\author{Jiahao Xun}
\affiliation{%
  \institution{Zhejiang Unversity, China}
  \country{}}
\email{jhxun@zju.edu.cn}

\author{Shengyu Zhang}
\affiliation{%
  \institution{Zhejiang Unversity, China}
  \country{}}
\email{sy_zhang@zju.edu.cn}

\author{Zhou Zhao\textsuperscript{\Letter}}
\affiliation{%
  \institution{Zhejiang Unversity, China}
  \country{}}
\email{zhaozhou@zju.edu.cn}

\author{Liqun Deng}
\affiliation{%
  \institution{Huawei Noah’s Ark Lab, China}
  \country{}}
\email{dengliqun.deng@huawei.com}

\author{Wenqiao Zhang}
\affiliation{%
  \institution{Zhejiang Unversity, China}
  \country{}}
\email{wenqiaozhang@zju.edu.cn}

\author{Zhenhua Dong}
\affiliation{%
  \institution{Huawei Noah’s Ark Lab, China}
  \country{}}
\email{dongzhenhua@huawei.com}

\author{Xin Jiang}
\affiliation{%
  \institution{Huawei Noah's Ark Lab, China}
  \country{}}
\email{Jiang.Xin@huawei.com}

\thanks{$^*$ Equal contribution.}
\thanks{\textsuperscript{\Letter} Corresponding author.}
\renewcommand{\authors}{Dong Yao, Jieming Zhu, Jiahao Xun, Shengyu Zhang, Liqun Deng, Zhou Zhao, Wenqiao Zhang, Zhenhua Dong, and Xin Jiang}
\renewcommand{\shortauthors}{Dong Yao et al.}

\newcommand{\zjm}[1]{\textcolor{magenta}{[#1]$_{zjm}$}}

\begin{abstract}
Recent research in self-supervised contrastive learning of music representations has demonstrated remarkable results across diverse downstream tasks. However, a prevailing trend in existing methods involves representing equally-sized music clips in either waveform or spectrogram formats, often overlooking the intrinsic part-whole hierarchies within music. In our quest to comprehend the bottom-up structure of music, we introduce MART, a hierarchical music representation learning approach that facilitates feature interactions among cropped music clips while considering their part-whole hierarchies. Specifically, we propose a hierarchical part-whole transformer to capture the structural relationships between music clips in a part-whole hierarchy. Furthermore, a hierarchical contrastive learning objective is crafted to align part-whole music representations at adjacent levels, progressively establishing a multi-hierarchy representation space. The effectiveness of our music representation learning from part-whole hierarchies has been empirically validated across multiple downstream tasks, including music classification and cover song identification.
\end{abstract}

\ccsdesc[500]{Computing methodologies~Artificial intelligence}

\keywords{Music representation learning, Part-whole hierarchy, Transformer}



\maketitle

\section{Introduction} \label{sec:intro}

Amid the remarkable success of self-supervised learning in diverse domains, self-supervised music representation learning has emerged as a prevalent approach, providing substantial advantages for various downstream tasks, such as music classification~\cite{Pons2018,Won2020} and cover song identification~\cite{Yu2020,Yu2019,DisCover}. Within the realm of music research, a variety of avenues have been explored, including: 1) initiatives aimed at extending the achievements of general audio representation~\cite{Niizumi2021, Saeed2021} into the music domain through the application of contrastive learning~\cite{Spijkervet2021} or masked autoencoders~\cite{MERT} techniques; 2) investigations into symbolic music understanding by framing music as MIDI-like sequences and leveraging pretrained language models, such as MusicBert~\cite{Zeng2021}; and 3) recent models that delve into language-music pretraining to align lyrics and audio~\cite{CLaMP} or generate music from text~\cite{MusicLM}.

In this paper, our primary focus lies on contrastive representation learning of music audios, given that, in many commercial settings, music lyrics and symbolic information (e.g., MIDI files) are not necessarily available. Despite significant progress made by previous work, a prevailing trend in existing methods involves cropping audios into equally-sized clips and learning to align clip representations from the same audio~\cite{Spijkervet2021, Niizumi2021, PEMR}, often overlooking the intrinsic part-whole hierarchical structures encoded in music. In musical form, individual notes and chords serve as the basic building blocks to form a measure. One measure or a few can come together to create a phrase. When several phrases are combined, a passage is formed, eventually leading to a movement~\cite{BerardinisVCC20}. This naturally results in a part-whole hierarchy crucial for music representation. For instance, cover song identification~\cite{Yu2020, Yu2019} requires assessing the similarity of two songs with different lengths, and multi-label music classification~\cite{Won2020} requires different music clips reflecting various labels. While previous attempts have explored long-term music structure for autoregressive music generation~\cite{MusicTransformer}, the question of how to effectively leverage part-whole hierarchies for music representation learning remains an open challenge.

\begin{figure*}[!th]
	\centering  
	\subfigure[Overall Architecture]
 {\label{fig:arch_over}
\includegraphics[width=0.55\linewidth]{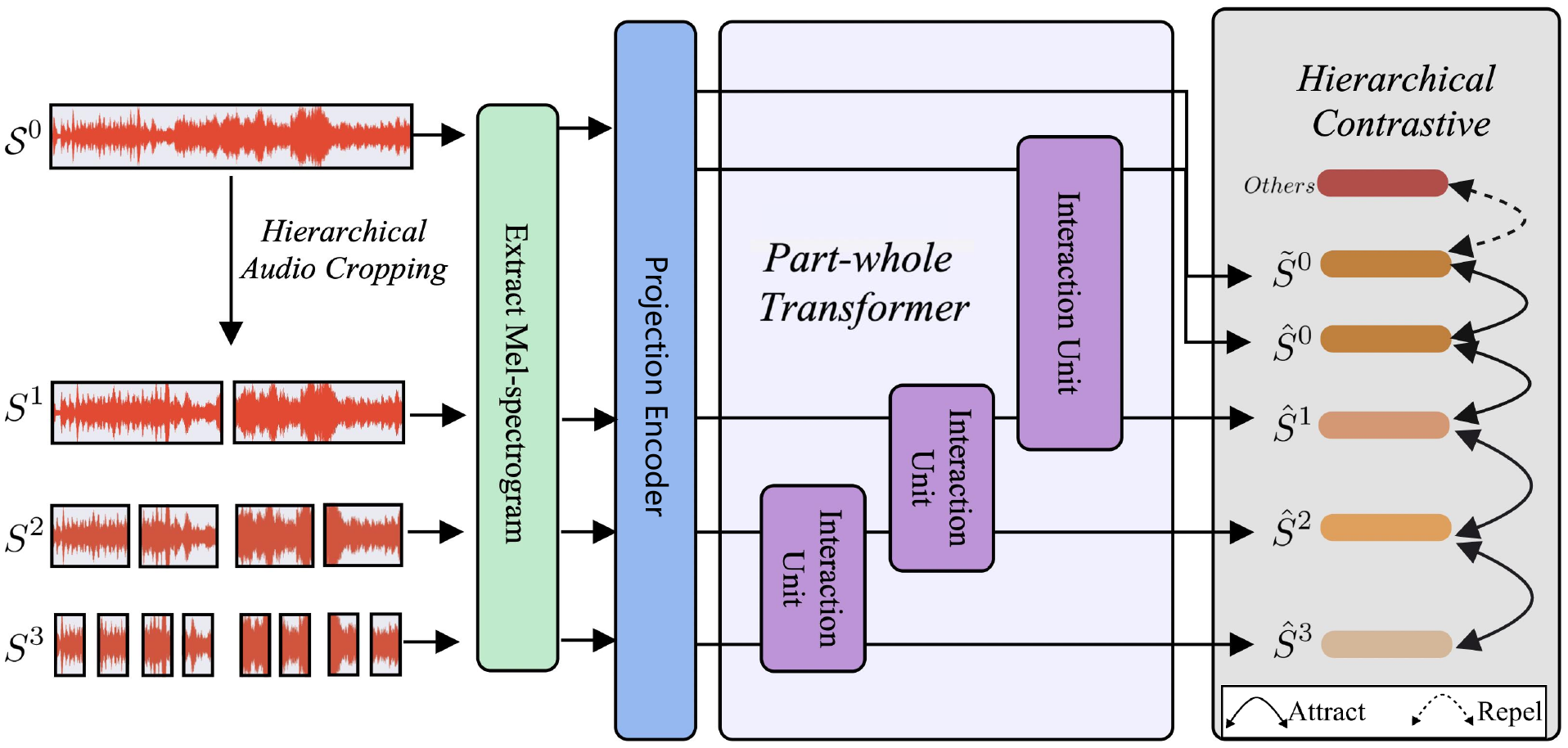}    
   } 
  \hspace{0.04\linewidth}
	\subfigure[Interaction Unit]
 {\label{fig:inter_unit}
\includegraphics[width=0.23\linewidth]{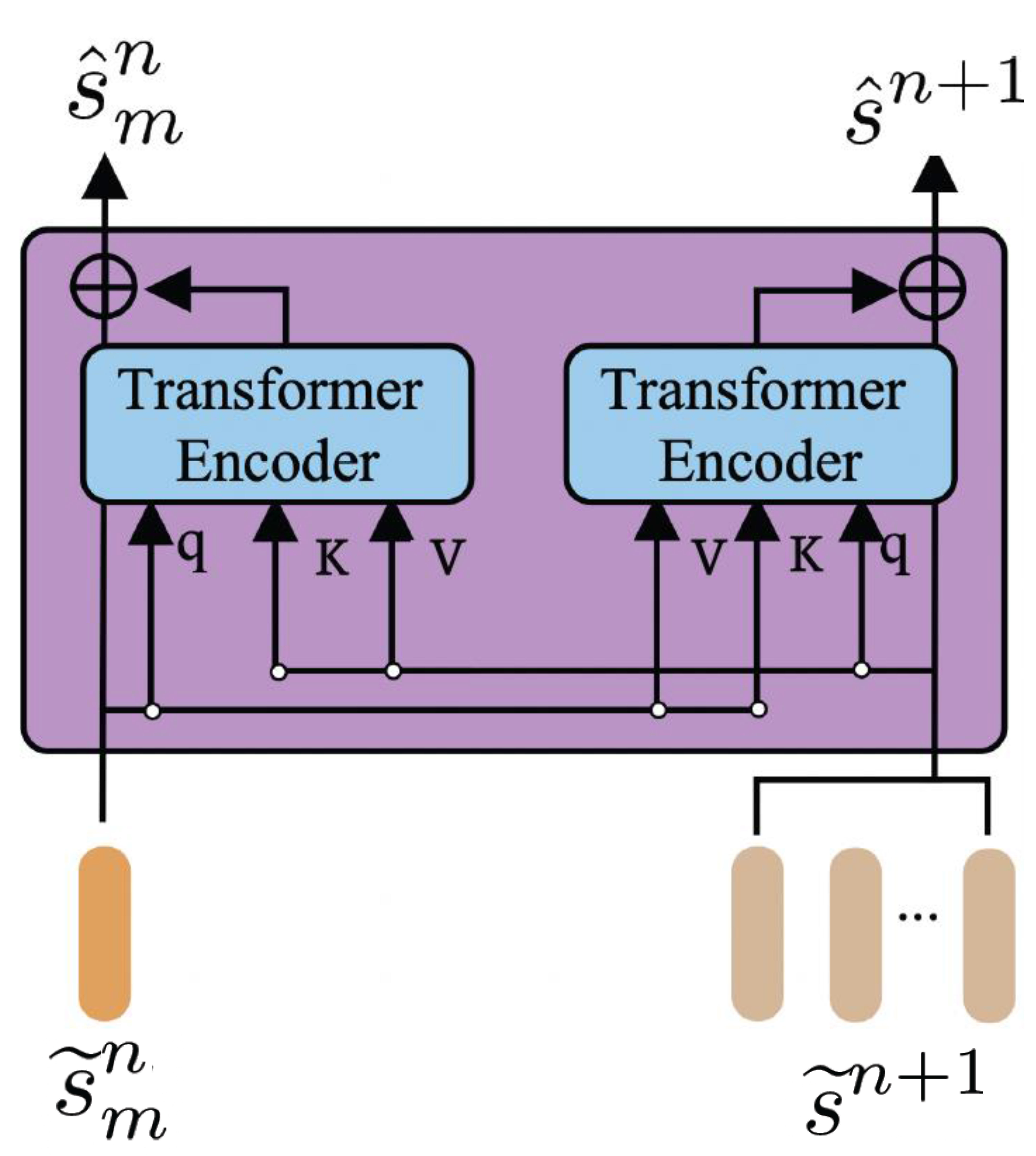}
 }
      \vspace{-2ex}
	\caption{The architecture of MART}
 \label{fig:fig1}
\end{figure*}

In our quest to comprehend the bottom-up structure of music, we propose a novel hierarchical part-whole contrastive learning approach named the Music Audio Representation Transformer (MART). MART facilitates feature interactions among cropped music clips while taking into consideration their part-whole hierarchies. Our work draws large inspiration from Hinton's conceptual GLOM framework~\cite{GLOM} on representing part-whole hierarchies in a neural network. 
Specifically, we propose a part-whole transformer to capture the structural relationships between music clips within a part-whole hierarchy. Additionally, we craft a hierarchical contrastive learning objective to align part-whole music representations at adjacent levels, thereby progressively establishing a multi-hierarchy representation space. We validate the effectiveness of our MART approach for music representation learning on both music classification and cover song identification tasks.

\section{Approach} \label{sec:method}






Our approach consists of three key modules: hierarchical audio cropping, part-whole transformer, and hierarchical contrastive loss. The overall architecture is shown in Figure \ref{fig:fig1}.



\subsection{Hierarchical Audio Cropping} \label{crop}
While the hierarchical structure is prevalent in music, extracting this structure remains an open problem. To enhance our hierarchical music representation learning, we propose a simple yet effective strategy termed Hierarchical Audio Cropping (HAC). This strategy involves recursively cropping a music audio clip into multiple sub-clips, thereby generating a diverse set of music clips with inherent part-whole hierarchies. Specifically, we divide the input music audio $S^0$ into $M$ (where $M$=2 in Figure \ref{fig:arch_over}) music clips of equal length, denoted as $S^1$. Then, we further split each music clip from $S^1$ into $M$ shorter music clips, i.e., $S^2$. This process repeats until we obtain $N$ hierarchical levels of music clips (where $N$=4 in Figure \ref{fig:arch_over}). Our hierarchical audio cropping strategy ensures that each short (part) music clip is a component of a long (whole) clip, resulting in a cohesive part-whole hierarchy. For clarity, we designate music clips at two adjacent levels as "part clips" and "whole clips", respectively.

\subsection{Part-Whole Transformer}\label{inter}

Upon acquiring music clips, we extract mel-spectrograms and transform them into the logarithmic scale, adhering to the common practice outlined in \cite{Saeed2021}. To accommodate music clips of varying lengths, we adjust hop lengths to derive log-scale mel-spectrograms of equal size. Subsequently, we employ a CNN-based projection encoder to project the mel-spectrograms into a shared latent space with consistent dimensionality. Formally, let $S^n = \{s^n_m | n \in [0, N), m \in [0, M^{n} ) \}$ denote the set of music clips obtained at the $n$-th level through hierarchical audio cropping, where $s^n_{m}$ represents the $m$-th music clip in the set $S^n$. In Figure~\ref{fig:arch_over}, the size of $S^3$ equals $M^3=8$. We utilize a projection encoder $E(s^n_m)$ to project each music clip into a hidden vector $\widetilde{s}^n_m$.


\subsubsection{Part-Whole Interaction Unit} 
Existing encoders often fall short in capturing the part-whole structure in music. In response, our work introduces a part-whole interaction module designed to facilitate interactions between part clips and whole clips, enabling the model to learn from their inherent part-whole relationships. As depicted in Figure \ref{fig:inter_unit}, we employ a part-whole pair as an illustrative example to explain this module. Specifically, we choose a whole clip $s^{n}_m$ from $S^n$ and its part clips ${s}^{n+1} = \{s^{n+1}_m | m \in [0, M )\}$ from $S^{n+1}$. After encoding by the projection encoder, we obtain hidden vectors $\widetilde{s}^n_m \in \mathbb{R}^{1 \times D_e}$ and $\widetilde{s}^{n+1} \in \mathbb{R}^{M \times D_e}$, respectively, where $D_e$ denotes the feature dimension after the projection encoder. Following the \textit{query}, \textit{key}, \textit{value} transformations typical of transformers, we derive  $Q^n, K^n, V^n \in \mathbb{R}^{1 \times D_t}$ by applying linear transformations to $\widetilde{s}^n_m$. Similarly, we obtain $Q^{n+1}, K^{n+1}, V^{n+1} \in \mathbb{R}^{M \times D_t}$ by transforming $\widetilde{s}^{n+1}$. To consolidate information from part $\widetilde{s}^{n+1}$ to whole $\widetilde{s}^n_m$, we employ  $Q^n, K^{n+1}, V^{n+1}$ as \textit{Query}, \textit{Key}, \textit{Value} in the transformer encoder. Subsequently, we derive the transformer output of the whole clip as follows:
\begin{equation} \label{eq:long}
    \hat{s}^n_m = \widetilde{s}^n_m + \lambda^n Transformer(Q^n, K^{n+1}, V^{n+1})
\end{equation}
where $\lambda^n$ is a hyperparameter that governs the significance of the interacted result, and $Transformer(\cdot)$ denotes the original transformer encoder. Symbols with tildes ($\widetilde{s}$) and hats ($\hat{s}$) above represent the input and output vectors, respectively. Consequently, the whole clip functions as a query vector, encoding key and value vectors of part clips through the multi-head attention mechanism. Conversely, when distilling the whole information $\widetilde{s}^n_m$ into part representations $\widetilde{s}^{n+1}$, employ the subsequent transformer encoder with $Q^{n+1}, K^n, V^n$ as inputs:
\begin{equation} \label{eq:short}
    \hat{s}^{n+1} = \widetilde{s}^{n+1} + \lambda^{n+1} Transformer(Q^{n+1}, K^n, V^n)
\end{equation}
where $\lambda^{n+1}$  represents a hyperparameter aimed at balancing the importance of the interacted result. In this scenario, the part clips function as query vectors, collecting global information from the whole clip. It is noteworthy that while multiple part-whole clip pairs exist at adjacent levels (e.g., 4 between $S^2$ and $S^3$), their interactions can be computed in parallel through an interaction unit. This resembles the multi-head mechanism, as each pair can be transformed into a separate head subspace.

\subsubsection{Part-Whole Transformer} 
The part-whole interaction unit stands as a fundamental component within the part-whole transformer. The part-whole interaction unit stands as a fundamental component within the part-whole transformer. Through the stacking of multiple interaction units, we can systematically propagate information and facilitate their interactions across the part-whole hierarchies. Note that music clips in the intermediate levels (e.g., $S^1$, $S^2$) serve as both whole clips in a lower interaction unit and part clips in an upper interaction unit. Simultaneously, we have the flexibility to stack multiple part-whole transformers as original transformer blocks, thereby enhancing the capacity and interaction ability of our model.

\subsection{Hierarchical Contrastive Learning}\label{hcl}
To utilize the information from part-whole interactions in the last step, we design a hierarchical contrastive learning method aimed at learning part and whole representations. Following established contrastive learning methods \cite{Grill2020}, we employ an MLP to project the outputs of part-whole transformers into contrastive latent vectors. In order to align part-whole clip representations at adjacent levels, for each data instance in the dataset, we create pairs of various hierarchical part-whole pairs, denoted as $P = \{(\hat{S}^0, \hat{S}^1), (\hat{S}^1, \hat{S}^2), ..., (\hat{S}^{n}, \hat{S}^{n+1}), ...\}$, where $\hat{S}^{n} = \{\hat{s}^n_m | m \in [0, M^{n-1} ) \}$. Consequently, we can calculate the contrastive loss for different part-whole pairs of music clips:
\begin{align}
\mathcal{L}^{pw}_{b} & = \sum_{(\hat{S}^i, \hat{S}^{i+1})}^{P} \lambda^i \Big( \sum_{\hat{s}_{b}^{i}}^{\hat{S}^i} \sum_{\hat{s}_{b}^{i+1}}^{\hat{U}_i} \exp\big(sim(\hat{s}^{i}_{b}, \hat{s}^{i+1}_{b}) / \tau \big) \Big) \\
\lambda_i & = len(\hat{s}^{i+1}_b) ~/~ len(\hat{s}^i_b)
\end{align}
where $i$ signifies the $i$-th level in the part-whole hierarchy. $\hat{U}_i$ is a set contains sub-clips cropped from $\hat{s}^{i}_b$. $\lambda^i$ serves as a weight to balance the importance of the corresponding part-whole pair, and $len(\cdot)$ is a function that measures the timespan of the corresponding music clip, meaning that the greater the difference in length between two music clips, the less impact their loss term has. $sim(\cdot)$ calculates the cosine similarity between two vectors, and $\tau$ represents a temperature parameter. The subscript $b$ of $\mathcal{L}^{pw}_{b}$ denotes the instance index in the current batch. As depicted in Figure~\ref{fig:arch_over}, $\mathcal{L}^{pw}_{b}$ is computed over paired adjacent clips, making it distinct from the existing literature.

Additionally, we utilize the widely adopted contrastive loss~\cite{Spijkervet2021} to draw positive views (i.e., $\hat{s}^0_b$ and $\widetilde{s}^0_b$) closer and push away other negative views within the batch. Finally, we derive the hierarchical contrastive loss $\mathcal{L}^{hc}_b$ by integrating both loss terms as follows:
\begin{align}
     \mathcal{L}^{hc}_b &= - \log \frac{\mathcal{L}^{pw}_b + \exp\big(sim(\hat{s}^0_b, \widetilde{s}^0_b) / \tau \big)}{\mathcal{L}^{pw}_b + \mathcal{L}^{neg} }\\
     \mathcal{L}^{neg} &= \sum_{u=1}^B  \exp\big(sim(\hat{s}^0_b, \widetilde{s}^0_u) / \tau \big) + \sum_{u=1}^{B} \mathbbm{1}_{u \neq b} \exp \big(sim(\hat{s}^0_b, \hat{s}^0_u) / \tau \big)
\end{align}
where $\mathbbm{1}_{u \neq b} \in \{0, 1\}$ represents an indicator function that evaluates to 1 if $u \neq b$, and $B$ denotes the batch size. In practice, our $\mathcal{L}^{pw}_b$ functions as regularization terms for the traditional contrastive loss, contributing to the refinement of both part representations and whole representations in the part-whole hierarchy.


\section{Experiments} \label{sec:exper}

\subsection{Experimental Settings}

\noindent \textbf{Setup.} 
We evaluate the pretrained models on music classification (MTAT \cite{Law2009}, GTZAN~\cite{Genussov2010}) and cover song identification (SHS100K \cite{Xu2018}, Covers80 \cite{Xu2018}) tasks.
Note that we use the available subset of SHS100K (denoted as SHS100K-SUB), 
since many audio files cannot be crawled due to invalid copyrights. SHS100K-SUB contains 1,088 songs and all their cover versions, totaling 12,000 audio files in all. Following the previous works \cite{Yu2020, Yu2019}, we use ROC-AUC and PR-AUC as evaluation metrics for music classification and use MAP, Precision@10, and MR1 as indicators of cover song identification.
\noindent \textbf{Implementation.}
We utilize torchaudio for mel-spectrogram extraction, employing a 256-point FFT for STFT alongside 128 mel-bands. Subsequently, we transform the mel-spectrogram to the logarithmic scale following COLA \cite{Saeed2021}. Music clips of varying lengths are compressed to identical-size log-mel spectrograms through different hop lengths, facilitating processing by a single encoder. We employ FCN-7 as the spectrogram projection encoder and stack three part-whole transformers, each with three attention heads. An output projection head maps the 512-dimension representation to a contrastive space, with a layer configuration of Linear(512)-ReLU-Linear(256). The MART model is pretrained on the training data from MTAT and further evaluated on downstream tasks. We apply the same set of data augmentations as CLMR \cite{Spijkervet2021}, encompassing polarity inversion, noise, gain, filter, delay, and pitch shift. We use the Adam optimizer with a learning rate of 0.0003 and weight decay of $1.0\times 10^{-6}$ during pre-training. We set the batch size to 48 and train the model for 300 epochs, consuming approximately 100 hours on 2 RTX3090 GPUs. Early stopping is utilized for linear probing and finetuning experiments.


\subsection{Downstream Task Evaluation}



\begin{table}[!b]
\vspace{-1ex}
\centering
\caption{Music classification evaluation results on MTAT.}
\vspace{-1ex}
\resizebox{\linewidth}{!}{
\begin{tabular}{l|cc|cc|cc}
\toprule
\multirow{2}{*}{Methods} & \multicolumn{2}{c|}{Linear Probing} & \multicolumn{2}{c|}{Finetuning (1\%)} & \multicolumn{2}{c}{Finetuning (10\%)} \\ \cmidrule {2-7} 
 & ROC-AUC & PR-AUC & ROC-AUC & PR-AUC & ROC-AUC & PR-AUC \\ \midrule
CLMR~\cite{Won2020a} & 89.3 & 36.0 & 78.5 & 25.0 & 86.8 & 32.6 \\
Multi-Format~\cite{wangMultiFormatContrastiveLearning2021} & 87.0 & 31.3 & 78.7 & 23.5 & 87.2 & 32.3 \\
COLA~\cite{Saeed2021} & 89.1 & 36.0 & 80.5 & 26.6 & 87.4 & 32.7 \\
BYOL-A~\cite{Niizumi2021} & 89.2 & 36.3 & 79.7 & 26.3 & 87.6 & 32.9 \\
\textbf{MART} & \textbf{89.5} & \textbf{36.7} & \textbf{81.2} & \textbf{27.0} & \textbf{88.0} & \textbf{34.1} \\ \bottomrule
\end{tabular} \label{tab:mtat}
}
\vspace{-3ex}
\end{table}

\noindent \textbf{Music classification} aims to predict relevant tags for a given song. We assess MART on both GTZAN~\cite{Genussov2010} and MTAT~\cite{Law2009} datasets. Specifically, MART is pretrained on the training set of MTAT. The results presented in Table~\ref{tab:gtzan} showcase MART's performance when finetuned on the GTZAN data for music classification. MART outperforms both supervised and finetuning methods, indicating superior performance in cross-dataset transfer scenarios. Additionally, MART is evaluated on the MTAT dataset through linear probing (with a frozen encoder) and finetuning with small-scale labeled data (using 1\% and 10\% of training instances), designed to assess representation quality. The results in Table~\ref{tab:mtat} also highlight the effectiveness of MART, surpassing other existing contrastive audio representation methods.

\begin{table}[t]
 \caption{Music classification evaluation results on GTZAN.} \vspace{-1ex}
    \centering
    \resizebox{.65\linewidth}{!}{
         \begin{tabular}{l|cc}
        \toprule
        Method  & ROC-AUC & PR-AUC \\ 
        \midrule
        \textit{Supervised}:     \\
        FCN-4 \cite{Choi2016}   & 91.5 & 62.5  \\
        Musicnn \cite{Pons2018}   & 91.8 & 66.3 \\
        SampleCNN \cite{Lee2019}   & 90.7 & 63.8 \\
        SampleCNN+SE \cite{Kim2018}   & 90.1 & 61.5\\
        Self-Attention \cite{Won2019}  & 86.7 & 54.2 \\
        Harmonic CNN \cite{Won2020a}  & 93.5 & 70.9 \\
        Short-Chunk CNN \cite{Won2020}  & 94.1 & 71.0\\
        \midrule
        \textit{Finetuning}: \\
        Multi-Format~\cite{wangMultiFormatContrastiveLearning2021}  & 93.2 & 71.2 \\ 
        BYOL-A~\cite{Niizumi2021}  & ${94.6}$ & ${74.8}$ \\
        COLA~\cite{Saeed2021}  & 94.2 & 72.6 \\
        
        \textbf{MART} & \textbf{94.7} & \textbf{77.2} \\
        \bottomrule
        \end{tabular}}
   
    \label{tab:gtzan}
    \vspace{-1ex}
\end{table}
\begin{table}[h]
\caption{Cover song identification evaluation results.} \vspace{-1ex}
    \centering
    \resizebox{\linewidth}{!}{
    \begin{tabular} {l|ccc|ccc}
    \toprule
    \multirow{2}{*}{Method} & \multicolumn{3}{c}{SHS100K-SUB}  & \multicolumn{3}{c}{Covers80} \\
    & MAP $\uparrow$ & P@10 $\uparrow$ & MR1 $\downarrow$ & MAP $\uparrow$ & P@10 $\uparrow$ & MR1 $\downarrow$\\
    \midrule
    Ki-Net \cite{Xu2018}  & 11.2 & 15.6 & 68.33 & 36.8 & 5.2 & 32.10\\
    TPP-Net \cite{Yu2019} & 26.7 & 21.7 & 35.75 & 50.0 & 6.8 & 17.08\\
    FCN-4  \cite{Choi2016}   & 28.9 & 23.0 & 34.86 & 52.9 & 7.3 & 12.50\\
    CQT-Net \cite{Yu2020} & 44.6 & 32.3 & 18.09 & 66.6 & 7.7 & 12.20\\
    \midrule
    BYOL-A~\cite{Niizumi2021} & 46.2 & 33.4 & 19.78 & 73.0 & 8.4 & \textbf{6.67}\\ 
    Multi-Format~\cite{wangMultiFormatContrastiveLearning2021}  & 47.7 & 33.9 & 19.74 & 71.8 & 8.3 & 8.83\\
    COLA~\cite{Saeed2021} & 47.2 & 34.0 & 20.15 & 74.6 & 8.5 & 7.86\\
    \textbf{MART}  & \textbf{52.2} & \textbf{35.8} & \textbf{14.53} & \textbf{74.8} & \textbf{8.6} & {7.53} \\
    \bottomrule
    \end{tabular}}

\label{tab:trans_cover}
\vspace{-1ex}
\end{table}
\noindent \textbf{Cover song identification} \cite{Yu2019, Yu2020} aims to recognize an alternative version of a given music file within a corpus. A cover song typically refers to a new performance or recording by a musician other than the original performer or composer of the song. Table \ref{tab:trans_cover} presents our evaluation results on SHS100K-SUB and Covers80 datasets. The upper part displays the results of supervised learning methods, while the lower part showcases our pretrained methods. Specifically, we incorporate pretrained encoders on top of CQTNet and finetune the combined network using SHS100K-SUB's training set, subsequently evaluating on SHS100K-SUB's test set and Covers80 (w.r.t. zero-shot transfer). The results highlight the benefits of pretrained methods, which significantly outperform the previous supervised methods. Remarkably, MART stands out among them, demonstrating noteworthy performance improvements on SHS100K-SUB. In the zero-shot setting, MART also achieves better or similar results on Covers80.

\begin{table}[t]
\centering
\caption{Ablation study on MTAT.} \vspace{-1ex}
\resizebox{0.9\linewidth}{!}{
\begin{tabular}{lcccc}
\toprule
Variant   & ROC-AUC  & RelaImpr. & PR-AUC   & RelaImpr. \\ \midrule
MART & 89.5 & -        & 36.7 & -        \\
w/o. PWT  & 89.2 & -0.32\%  & 36.5 & -0.54\%  \\
w/o. HCL  & 89.3 & -0.21\%  & 36.5 & -.0.44\% \\
w/o. PWT \& HCL  & 89.1 & -0.44\%  & 36.2 & -1.18\%  \\ \bottomrule
\end{tabular}}
\label{tab:abla}
\end{table}

\subsection{Ablation Study}\label{subsec:model_analysis}

We perform an ablation study on the MTAT dataset to assess the contributions of key components in MART, namely the part-whole transformer (PWT) and hierarchical contrastive learning (HCL). More specifically, we compare our model with three variants: without PWT, without HCL, and without both. The experiments are conducted via linear probing, and the results are presented in Table \ref{tab:abla}. In essence, the removal of any component results in a performance decline. The most substantial decrease occurs when both components are removed, signifying a complete loss of part-whole structure learning. These outcomes underscore the efficacy of MART in hierarchical part-whole representation learning.

\section{Conclusion} \label{sec:conclusion}
In this study, we introduce MART, a self-supervised music representation learning approach that leverages the intrinsic part-whole hierarchies within music for feature interactions, thereby constructing a multi-hierarchy latent space. To this end, we employ a hierarchical contrastive learning objective to align part-whole music representations within adjacent hierarchies. The effectiveness of MART has been demonstrated across various downstream tasks. Looking ahead, for future work, we aim to extend MART's training to massive open-domain music datasets, with the goal of further enhancing the model's zero-shot generability.

\balance
\bibliography{MART.bib}


\begin{thebibliography}{26}


\ifx \showCODEN    \undefined \def \showCODEN     #1{\unskip}     \fi
\ifx \showDOI      \undefined \def \showDOI       #1{#1}\fi
\ifx \showISBNx    \undefined \def \showISBNx     #1{\unskip}     \fi
\ifx \showISBNxiii \undefined \def \showISBNxiii  #1{\unskip}     \fi
\ifx \showISSN     \undefined \def \showISSN      #1{\unskip}     \fi
\ifx \showLCCN     \undefined \def \showLCCN      #1{\unskip}     \fi
\ifx \shownote     \undefined \def \shownote      #1{#1}          \fi
\ifx \showarticletitle \undefined \def \showarticletitle #1{#1}   \fi
\ifx \showURL      \undefined \def \showURL       {\relax}        \fi
\providecommand\bibfield[2]{#2}
\providecommand\bibinfo[2]{#2}
\providecommand\natexlab[1]{#1}
\providecommand\showeprint[2][]{arXiv:#2}

\bibitem[Agostinelli and et~al.(2023)]%
        {MusicLM}
\bibfield{author}{\bibinfo{person}{Andrea Agostinelli} {and} \bibinfo{person}{et al.}} \bibinfo{year}{2023}\natexlab{}.
\newblock \showarticletitle{MusicLM: Generating Music From Text}.
\newblock \bibinfo{journal}{\emph{CoRR}}  \bibinfo{volume}{abs/2301.11325} (\bibinfo{year}{2023}).
\newblock


\bibitem[Choi et~al\mbox{.}(2016)]%
        {Choi2016}
\bibfield{author}{\bibinfo{person}{Keunwoo Choi}, \bibinfo{person}{Gy{\"{o}}rgy Fazekas}, {and} \bibinfo{person}{Mark~B. Sandler}.} \bibinfo{year}{2016}\natexlab{}.
\newblock \showarticletitle{Automatic Tagging Using Deep Convolutional Neural Networks}. In \bibinfo{booktitle}{\emph{ISMIR}}. \bibinfo{pages}{805--811}.
\newblock


\bibitem[de~Berardinis et~al\mbox{.}(2020)]%
        {BerardinisVCC20}
\bibfield{author}{\bibinfo{person}{Jacopo de Berardinis}, \bibinfo{person}{Michalis Vamvakaris}, \bibinfo{person}{Angelo Cangelosi}, {and} \bibinfo{person}{Eduardo Coutinho}.} \bibinfo{year}{2020}\natexlab{}.
\newblock \showarticletitle{Unveiling the Hierarchical Structure of Music by Multi-Resolution Community Detection}.
\newblock \bibinfo{journal}{\emph{TISMIR}} \bibinfo{volume}{3}, \bibinfo{number}{1} (\bibinfo{year}{2020}), \bibinfo{pages}{82--97}.
\newblock


\bibitem[Genussov and Cohen(2010)]%
        {Genussov2010}
\bibfield{author}{\bibinfo{person}{Michal Genussov} {and} \bibinfo{person}{Israel Cohen}.} \bibinfo{year}{2010}\natexlab{}.
\newblock \showarticletitle{Musical genre classification of audio signals using geometric methods}. In \bibinfo{booktitle}{\emph{EUSIPCO}}. \bibinfo{pages}{497--501}.
\newblock


\bibitem[Grill et~al\mbox{.}(2020)]%
        {Grill2020}
\bibfield{author}{\bibinfo{person}{Jean-Bastien Grill}, \bibinfo{person}{Florian Strub}, \bibinfo{person}{Florent Altch{\'e}}, \bibinfo{person}{Corentin Tallec}, \bibinfo{person}{Pierre Richemond}, \bibinfo{person}{Elena Buchatskaya}, \bibinfo{person}{Carl Doersch}, \bibinfo{person}{Bernardo Avila~Pires}, \bibinfo{person}{Zhaohan Guo}, \bibinfo{person}{Mohammad Gheshlaghi~Azar}, {et~al\mbox{.}}} \bibinfo{year}{2020}\natexlab{}.
\newblock \showarticletitle{Bootstrap your own latent-a new approach to self-supervised learning}.
\newblock \bibinfo{journal}{\emph{NeurIPS}}  \bibinfo{volume}{33} (\bibinfo{year}{2020}), \bibinfo{pages}{21271--21284}.
\newblock


\bibitem[Hinton(2023)]%
        {GLOM}
\bibfield{author}{\bibinfo{person}{Geoffrey~E. Hinton}.} \bibinfo{year}{2023}\natexlab{}.
\newblock \showarticletitle{How to Represent Part-Whole Hierarchies in a Neural Network}.
\newblock \bibinfo{journal}{\emph{Neural Comput.}} \bibinfo{volume}{35}, \bibinfo{number}{3} (\bibinfo{year}{2023}), \bibinfo{pages}{413--452}.
\newblock


\bibitem[Huang et~al\mbox{.}(2019)]%
        {MusicTransformer}
\bibfield{author}{\bibinfo{person}{Cheng{-}Zhi~Anna Huang}, \bibinfo{person}{Ashish Vaswani}, \bibinfo{person}{Jakob Uszkoreit}, \bibinfo{person}{Ian Simon}, \bibinfo{person}{Curtis Hawthorne}, \bibinfo{person}{Noam Shazeer}, \bibinfo{person}{Andrew~M. Dai}, \bibinfo{person}{Matthew~D. Hoffman}, \bibinfo{person}{Monica Dinculescu}, {and} \bibinfo{person}{Douglas Eck}.} \bibinfo{year}{2019}\natexlab{}.
\newblock \showarticletitle{Music Transformer: Generating Music with Long-Term Structure}. In \bibinfo{booktitle}{\emph{ICLR}}.
\newblock


\bibitem[Kim et~al\mbox{.}(2018)]%
        {Kim2018}
\bibfield{author}{\bibinfo{person}{Taejun Kim}, \bibinfo{person}{Jongpil Lee}, {and} \bibinfo{person}{Juhan Nam}.} \bibinfo{year}{2018}\natexlab{}.
\newblock \showarticletitle{Sample-level CNN architectures for music auto-tagging using raw waveforms}. In \bibinfo{booktitle}{\emph{ICASSP}}. \bibinfo{pages}{366--370}.
\newblock


\bibitem[Law et~al\mbox{.}(2009)]%
        {Law2009}
\bibfield{author}{\bibinfo{person}{Edith Law}, \bibinfo{person}{Kris West}, \bibinfo{person}{Michael~I Mandel}, \bibinfo{person}{Mert Bay}, {and} \bibinfo{person}{J~Stephen Downie}.} \bibinfo{year}{2009}\natexlab{}.
\newblock \showarticletitle{Evaluation of algorithms using games: The case of music tagging.}. In \bibinfo{booktitle}{\emph{ISMIR}}.
\newblock


\bibitem[Lee et~al\mbox{.}(2019)]%
        {Lee2019}
\bibfield{author}{\bibinfo{person}{Jongpil Lee}, \bibinfo{person}{Jiyoung Park}, \bibinfo{person}{Keunhyoung~Luke Kim}, {and} \bibinfo{person}{Juhan Nam}.} \bibinfo{year}{2019}\natexlab{}.
\newblock \showarticletitle{{Sample-level deep convolutional neural networks for music auto-tagging using raw waveforms}}.
\newblock \bibinfo{journal}{\emph{SMC}} (\bibinfo{year}{2019}), \bibinfo{pages}{220--226}.
\newblock
\showISBNx{9789526037295}


\bibitem[Li et~al\mbox{.}(2023)]%
        {MERT}
\bibfield{author}{\bibinfo{person}{Yizhi Li}, \bibinfo{person}{Ruibin Yuan}, \bibinfo{person}{Ge Zhang}, \bibinfo{person}{Yinghao Ma}, \bibinfo{person}{Xingran Chen}, \bibinfo{person}{Hanzhi Yin}, \bibinfo{person}{Chenghua Lin}, \bibinfo{person}{Anton Ragni}, \bibinfo{person}{Emmanouil Benetos}, \bibinfo{person}{Norbert Gyenge}, \bibinfo{person}{Roger~B. Dannenberg}, \bibinfo{person}{Ruibo Liu}, \bibinfo{person}{Wenhu Chen}, \bibinfo{person}{Gus Xia}, \bibinfo{person}{Yemin Shi}, \bibinfo{person}{Wenhao Huang}, \bibinfo{person}{Yike Guo}, {and} \bibinfo{person}{Jie Fu}.} \bibinfo{year}{2023}\natexlab{}.
\newblock \showarticletitle{{MERT:} Acoustic Music Understanding Model with Large-Scale Self-supervised Training}.
\newblock \bibinfo{journal}{\emph{CoRR}}  \bibinfo{volume}{abs/2306.00107} (\bibinfo{year}{2023}).
\newblock


\bibitem[Niizumi et~al\mbox{.}(2021)]%
        {Niizumi2021}
\bibfield{author}{\bibinfo{person}{Daisuke Niizumi}, \bibinfo{person}{Daiki Takeuchi}, \bibinfo{person}{Yasunori Ohishi}, \bibinfo{person}{Noboru Harada}, {and} \bibinfo{person}{Kunio Kashino}.} \bibinfo{year}{2021}\natexlab{}.
\newblock \showarticletitle{Byol for audio: Self-supervised learning for general-purpose audio representation}. In \bibinfo{booktitle}{\emph{IJCNN}}. \bibinfo{pages}{1--8}.
\newblock


\bibitem[Pons~Puig et~al\mbox{.}(2018)]%
        {Pons2018}
\bibfield{author}{\bibinfo{person}{Jordi Pons~Puig}, \bibinfo{person}{Oriol Nieto~Caballero}, \bibinfo{person}{Matthew Prockup}, \bibinfo{person}{Erik~M Schmidt}, \bibinfo{person}{Andreas~F Ehmann}, {and} \bibinfo{person}{Xavier Serra}.} \bibinfo{year}{2018}\natexlab{}.
\newblock \showarticletitle{End-to-end learning for music audio tagging at scale}. In \bibinfo{booktitle}{\emph{ISMIR}}.
\newblock


\bibitem[Saeed et~al\mbox{.}(2021)]%
        {Saeed2021}
\bibfield{author}{\bibinfo{person}{Aaqib Saeed}, \bibinfo{person}{David Grangier}, {and} \bibinfo{person}{Neil Zeghidour}.} \bibinfo{year}{2021}\natexlab{}.
\newblock \showarticletitle{Contrastive Learning of General-Purpose Audio Representations}. In \bibinfo{booktitle}{\emph{ICASSP}}. \bibinfo{pages}{3875--3879}.
\newblock


\bibitem[Spijkervet and Burgoyne(2021)]%
        {Spijkervet2021}
\bibfield{author}{\bibinfo{person}{Janne Spijkervet} {and} \bibinfo{person}{John~Ashley Burgoyne}.} \bibinfo{year}{2021}\natexlab{}.
\newblock \showarticletitle{{Contrastive Learning of Musical Representations}}. In \bibinfo{booktitle}{\emph{ISMIR}}.
\newblock


\bibitem[Wang and van~den Oord(2021)]%
        {wangMultiFormatContrastiveLearning2021}
\bibfield{author}{\bibinfo{person}{Luyu Wang} {and} \bibinfo{person}{Aaron van~den Oord}.} \bibinfo{year}{2021}\natexlab{}.
\newblock \showarticletitle{Multi-{{Format Contrastive Learning}} of {{Audio Representations}}}.
\newblock  \bibinfo{number}{arXiv:2103.06508} (\bibinfo{date}{March} \bibinfo{year}{2021}).
\newblock


\bibitem[Won et~al\mbox{.}(2020a)]%
        {Won2020a}
\bibfield{author}{\bibinfo{person}{Minz Won}, \bibinfo{person}{Sanghyuk Chun}, \bibinfo{person}{Oriol Nieto}, {and} \bibinfo{person}{Xavier Serrc}.} \bibinfo{year}{2020}\natexlab{a}.
\newblock \showarticletitle{{Data-Driven Harmonic Filters for Audio Representation Learning}}.
\newblock \bibinfo{journal}{\emph{ICASSP}} (\bibinfo{year}{2020}), \bibinfo{pages}{536--540}.
\newblock
\showISBNx{9781509066315}
\showISSN{15206149}


\bibitem[Won et~al\mbox{.}(2019)]%
        {Won2019}
\bibfield{author}{\bibinfo{person}{Minz Won}, \bibinfo{person}{Sanghyuk Chun}, {and} \bibinfo{person}{Xavier Serra}.} \bibinfo{year}{2019}\natexlab{}.
\newblock \showarticletitle{{Toward Interpretable Music Tagging with Self-Attention}}.
\newblock  (\bibinfo{year}{2019}).
\newblock


\bibitem[Won et~al\mbox{.}(2020b)]%
        {Won2020}
\bibfield{author}{\bibinfo{person}{Minz Won}, \bibinfo{person}{Andres Ferraro}, \bibinfo{person}{Dmitry Bogdanov}, {and} \bibinfo{person}{Xavier Serra}.} \bibinfo{year}{2020}\natexlab{b}.
\newblock \showarticletitle{{Evaluation of CNN-based automatic music tagging models}}.
\newblock \bibinfo{journal}{\emph{SMC}} (\bibinfo{year}{2020}), \bibinfo{pages}{331--337}.
\newblock
\showISBNx{9788894541502}
\showISSN{25183672}


\bibitem[Wu et~al\mbox{.}(2023)]%
        {CLaMP}
\bibfield{author}{\bibinfo{person}{Shangda Wu}, \bibinfo{person}{Dingyao Yu}, \bibinfo{person}{Xu Tan}, {and} \bibinfo{person}{Maosong Sun}.} \bibinfo{year}{2023}\natexlab{}.
\newblock \showarticletitle{CLaMP: Contrastive Language-Music Pre-Training for Cross-Modal Symbolic Music Information Retrieval}. In \bibinfo{booktitle}{\emph{ISMIR}}. \bibinfo{pages}{157--165}.
\newblock


\bibitem[Xu et~al\mbox{.}(2018)]%
        {Xu2018}
\bibfield{author}{\bibinfo{person}{Xiaoshuo Xu}, \bibinfo{person}{Xiaoou Chen}, {and} \bibinfo{person}{Deshun Yang}.} \bibinfo{year}{2018}\natexlab{}.
\newblock \showarticletitle{Key-invariant convolutional neural network toward efficient cover song identification}. In \bibinfo{booktitle}{\emph{ICME}}. \bibinfo{pages}{1--6}.
\newblock


\bibitem[Xun et~al\mbox{.}(2023)]%
        {DisCover}
\bibfield{author}{\bibinfo{person}{Jiahao Xun}, \bibinfo{person}{Shengyu Zhang}, \bibinfo{person}{Yanting Yang}, \bibinfo{person}{Jieming Zhu}, \bibinfo{person}{Liqun Deng}, \bibinfo{person}{Zhou Zhao}, \bibinfo{person}{Zhenhua Dong}, \bibinfo{person}{Ruiqi Li}, \bibinfo{person}{Lichao Zhang}, {and} \bibinfo{person}{Fei Wu}.} \bibinfo{year}{2023}\natexlab{}.
\newblock \showarticletitle{DisCover: Disentangled Music Representation Learning for Cover Song Identification}. In \bibinfo{booktitle}{\emph{SIGIR}}. \bibinfo{pages}{453--463}.
\newblock


\bibitem[Yao et~al\mbox{.}(2022)]%
        {PEMR}
\bibfield{author}{\bibinfo{person}{Dong Yao}, \bibinfo{person}{Zhou Zhao}, \bibinfo{person}{Shengyu Zhang}, \bibinfo{person}{Jieming Zhu}, \bibinfo{person}{Yudong Zhu}, \bibinfo{person}{Rui Zhang}, {and} \bibinfo{person}{Xiuqiang He}.} \bibinfo{year}{2022}\natexlab{}.
\newblock \showarticletitle{Contrastive Learning with Positive-Negative Frame Mask for Music Representation}. In \bibinfo{booktitle}{\emph{{WWW}}}. \bibinfo{pages}{2906--2915}.
\newblock


\bibitem[Yu et~al\mbox{.}(2019)]%
        {Yu2019}
\bibfield{author}{\bibinfo{person}{Zhesong Yu}, \bibinfo{person}{Xiaoshuo Xu}, \bibinfo{person}{Xiaoou Chen}, {and} \bibinfo{person}{Deshun Yang}.} \bibinfo{year}{2019}\natexlab{}.
\newblock \showarticletitle{Temporal Pyramid Pooling Convolutional Neural Network for Cover Song Identification.}. In \bibinfo{booktitle}{\emph{IJCAI}}. \bibinfo{pages}{4846--4852}.
\newblock


\bibitem[Yu et~al\mbox{.}(2020)]%
        {Yu2020}
\bibfield{author}{\bibinfo{person}{Zhesong Yu}, \bibinfo{person}{Xiaoshuo Xu}, \bibinfo{person}{Xiaoou Chen}, {and} \bibinfo{person}{Deshun Yang}.} \bibinfo{year}{2020}\natexlab{}.
\newblock \showarticletitle{Learning a representation for cover song identification using convolutional neural network}. In \bibinfo{booktitle}{\emph{ICASSP}}. \bibinfo{pages}{541--545}.
\newblock


\bibitem[Zeng et~al\mbox{.}(2021)]%
        {Zeng2021}
\bibfield{author}{\bibinfo{person}{Mingliang Zeng}, \bibinfo{person}{Xu Tan}, \bibinfo{person}{Rui Wang}, \bibinfo{person}{Zeqian Ju}, \bibinfo{person}{Tao Qin}, {and} \bibinfo{person}{Tie-Yan Liu}.} \bibinfo{year}{2021}\natexlab{}.
\newblock \showarticletitle{{MusicBERT: Symbolic Music Understanding with Large-Scale Pre-Training}}.
\newblock  (\bibinfo{year}{2021}).
\newblock


\end{thebibliography}
\bibliographystyle{ACM-Reference-Format}

\clearpage
\appendix


\end{document}